\documentclass[a4paper,11pt]{article}

\usepackage{hyperref}%
\hypersetup{colorlinks=true,%
  linkcolor=blue,%
  citecolor=magenta,%
  urlcolor=black,%
  bookmarksnumbered=true,%
  bookmarkstype=toc,%
  bookmarksopen=false,%
  pdftitle={Algebraic Description of Shape Invariance Revisited},%
  pdfkeywords={shape invariance; representation theory},%
  pdfauthor={Satoshi Ohya}}

\usepackage[margin=1in]{geometry}

\usepackage[bitstream-charter,cal=cmcal]{mathdesign}
\usepackage[T1]{fontenc}

\usepackage{xcolor}

\usepackage{amsmath,amsbsy,bm}

\usepackage{epic,eepic}

\usepackage{tikz}
\usetikzlibrary{decorations.pathmorphing,matrix,calc}

\usepackage[hang,font=small,labelfont=bf]{caption}
\usepackage{subfigure}

\makeatletter%

\@addtoreset{equation}{section}%
\makeatother

\usepackage{titlesec}
\titleformat{\section}[block]{\filright\bfseries}{\thesection.}{0.5em}{}[]
\titleformat{\subsection}[block]{\filright\bfseries}{\thesubsection.}{0.5em}{}

\usepackage{titletoc}%
\contentsmargin{0cm}%
\titlecontents{section}[.8cm]{\filright\bfseries\boldmath}{\contentslabel{.8cm}}{}{\hfill\thecontentspage}[]
\titlecontents{subsection}[1.6cm]{\filright\small}{\contentslabel{.8cm}}{}{\;\titlerule*[.5pc]{.}\;\thecontentspage}[]
\titlecontents{subsubsection}[2.4cm]{\filright\small}{\contentslabel{.8cm}}{}{\;\titlerule*[.5pc]{.}\;\thecontentspage}[]

\usepackage{cite}

\title{\large\bfseries Algebraic Description of Shape Invariance
  Revisited}
\author{
  {\normalsize Satoshi Ohya}\\[1em]
  {\small\itshape Institute of Quantum Science, Nihon University}\\
  {\small\itshape Kanda-Surugadai 1-8-14, Chiyoda, Tokyo 101-8308,
    Japan}\\[1ex]
  {\small\ttfamily ohya@phys.cst.nihon-u.ac.jp}}
\date{\small (Dated: \today)}

\begin{document}
\maketitle
\flushbottom

\begin{abstract}
  We revisit the algebraic description of shape invariance method in
  one-dimensional quantum mechanics. In this note we focus on four
  particular examples: the Kepler problem in flat space, the Kepler
  problem in spherical space, the Kepler problem in hyperbolic space,
  and the Rosen--Morse potential problem. Following the prescription
  given by Gangopadhyaya \textit{et al.}, we first introduce certain
  nonlinear algebraic systems. We then show that, if the model
  parameters are appropriately quantized, the bound-state problems can
  be solved solely by means of representation theory.
\end{abstract}

\begingroup%
\hypersetup{linkcolor=black}
\tableofcontents
\endgroup

\newpage
\section{Introduction}
\label{section:1}
\begin{figure}[t]
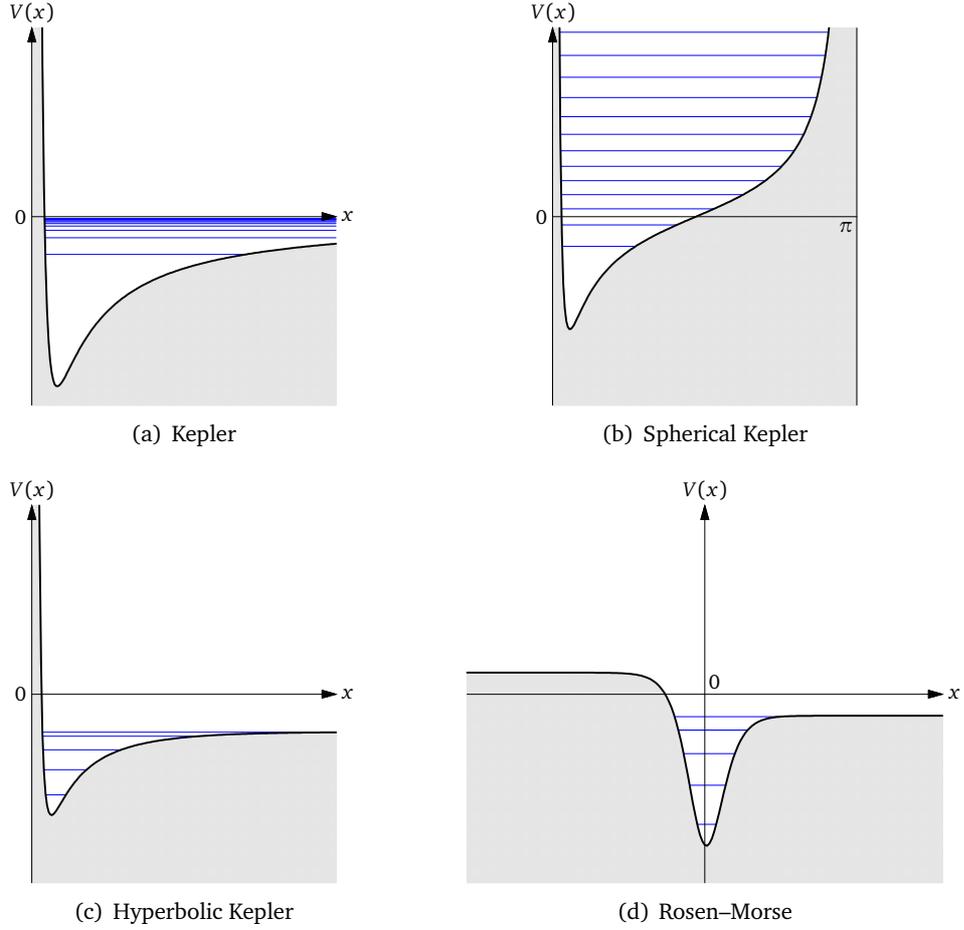

  \centering%
  \subfigure[Kepler]{\input{figure1a.eepic}}
  \subfigure[Spherical Kepler]{\input{figure1b.eepic}}\\
  \subfigure[Hyperbolic Kepler]{\input{figure1c.eepic}}
  \subfigure[Rosen--Morse]{\input{figure1d.eepic}}
  \caption{Potential energies (thick solid curves) and discrete energy
    levels (blue lines).}
  \label{figure:1}
\end{figure}

The purpose of this note is to revisit a couple of one-dimensional
quantum-mechanical bound-state problems that can be solved exactly. In
this note we shall focus on four particular examples: the Kepler
problem in flat space, the Kepler problem in spherical space
\cite{Schrodinger:1940xj,Infeld:1941,Stevenson:1941}, the Kepler
problem in hyperbolic space \cite{Manning:1933,Infeld:1945}, and the
Rosen--Morse potential problem \cite{Eckart:1930zza,Rosen:1932}, all
of whose bound-state spectra are known to be exactly
calculable. Hamiltonians of these problems\footnote{These names for
  the Hamiltonians, though not so popular nowadays, are borrowed (with
  slight modifications) from Infeld and Hull
  \cite{Infeld:1951mw}. Notice that these are different from those
  commonly used in the supersymmetric quantum mechanics literature
  \cite{Cooper:1994eh}.} are respectively given by
\begin{subequations}
  \begin{align}
    H_{\text{Kepler}}&=-\frac{d^{2}}{dx^{2}}+\frac{j(j-1)}{x^{2}}-\frac{2g}{x},\label{eq:1.1a}\\
    H_{\text{spherical Kepler}}&=-\frac{d^{2}}{dx^{2}}+\frac{j(j-1)}{\sin^{2}x}-2g\cot x,\label{eq:1.1b}\\
    H_{\text{hyperbolic Kepler}}&=-\frac{d^{2}}{dx^{2}}+\frac{j(j-1)}{\sinh^{2}x}-2g\coth x,\label{eq:1.1c}\\
    H_{\text{Rosen--Morse}}&=-\frac{d^{2}}{dx^{2}}-\frac{j(j-1)}{\cosh^{2}x}-2g\tanh x,\label{eq:1.1d}
  \end{align}
\end{subequations}
where $j$ and $g$ are real parameters. The potential energies and
bound-state spectra are depicted in Figure \ref{figure:1}.

There exist several methods to solve the eigenvalue problems of these
Hamiltonians \eqref{eq:1.1a}--\eqref{eq:1.1d}. Among them is the shape
invariance method \cite{Cooper:1994eh},\footnote{Recently it has been
  demonstrated that spectral intertwining relation provides a yet
  another scheme to solve the eigenvalue problems of
  \eqref{eq:1.1a}--\eqref{eq:1.1c} \cite{Houri:2017xtq}.} which is
based on the factorization of Hamiltonian and the Darboux
transformation. And, as discussed by Gangopadhyaya \textit{et al.}
\cite{Gangopadhyaya:1998ccj} (see also the reviews
\cite{Rasinariu:2007,Bougie:2012}), the shape invariance can always be
translated into the (Lie-)algebraic description---the so-called
potential algebra.\footnote{A similar algebraic description for shape
  invariance has also been discussed by Balantekin
  \cite{Balantekin:1997mg}.} The spectral problem can then be solved
by means of representation theory. However, as far as we noticed, the
representation theory of potential algebra has not been fully analyzed
yet. In particular, the spectral problems of the above Hamiltonians
have not been solved in terms of potential algebra. The purpose of
this note is to fill this gap. As we will see below, these very old
spectral problems require to introduce rather nontrivial nonlinear
algebraic systems. The goal of this note is to show that these
bound-state problems can be solved by representation theory of the
operators $\{J_{3},J_{+},J_{-}\}$ that satisfy the linear commutation
relations between $J_{3}$ and $J_{\pm}$
\begin{align}
  [J_{3},J_{\pm}]=\pm J_{\pm},\label{eq:1.2}
\end{align}
and the nonlinear commutation relations between $J_{+}$ and $J_{-}$
\begin{subequations}
  \begin{align}
    \text{(Kepler)}\quad[J_{+},J_{-}]&=-\frac{g^{2}}{J_{3}^{2}}+\frac{g^{2}}{(J_{3}-1)^{2}},\label{eq:1.3a}\\
    \text{(spherical Kepler)}\quad[J_{+},J_{-}]&=J_{3}^{2}-\frac{g^{2}}{J_{3}^{2}}-(J_{3}-1)^{2}+\frac{g^{2}}{(J_{3}-1)^{2}},\label{eq:1.3b}\\
    \text{(hyperbolic Kepler \& Rosen--Morse)}\quad[J_{+},J_{-}]&=-J_{3}^{2}-\frac{g^{2}}{J_{3}^{2}}+(J_{3}-1)^{2}+\frac{g^{2}}{(J_{3}-1)^{2}}.\label{eq:1.3c}
  \end{align}
\end{subequations}
We will see that, if $j$ is a half-integer, the bound-state problems
of \eqref{eq:1.1a}--\eqref{eq:1.1d} can be solved from these
operators.

The rest of the note is organized as follows: In Section
\ref{section:2} we introduce the potential algebra for the Kepler
problem in flat space and solve the spectral problem by means of
representation theory. In Sections \ref{section:3} and \ref{section:4}
we generalize to the other problems. We shall see that the bound-state
spectra of the hyperbolic Kepler and Rosen--Morse Hamiltonians just
correspond to two distinct representations of the same algebraic
system. We conclude in Section \ref{section:5}.

\section{Kepler}
\label{section:2}
Let us start with the Kepler problem in flat space. As is well known,
the Kepler Hamiltonian \eqref{eq:1.1a} can be factorized as follows:
\begin{align}
  H_{\text{Kepler}}=A_{-}(j)A_{+}(j)-\frac{g^{2}}{j^{2}},\label{eq:2.1}
\end{align}
where $A_{\pm}(j)$ are the first-order differential operators given by
\begin{align}
  A_{\pm}(j)=\pm\frac{d}{dx}-\frac{j}{x}+\frac{g}{j}.\label{eq:2.2}
\end{align}
Let us next introduce the potential algebra of this system. Following
Ref.~\cite{Gangopadhyaya:1998ccj} with slight modifications, we first
introduce an auxiliary periodic variable $\theta\in[0,2\pi)$, then
upgrade the parameter $j$ to an operator $J_{3}=-i\partial_{\theta}$,
and then replace $A_{+}(j)$ and $A_{-}(j)$ to
$J_{+}=\mathrm{e}^{i\theta}A_{+}(J_{3})$ and
$J_{-}=A_{-}(J_{3})\mathrm{e}^{-i\theta}$. The resultant operators
that we wish to study are thus as follows:
\begin{subequations}
  \begin{align}
    J_{3}&=-i\partial_{\theta},\label{eq:2.3a}\\
    J_{+}&=\mathrm{e}^{i\theta}\left(\partial_{x}-\frac{J_{3}}{x}+\frac{g}{J_{3}}\right),\label{eq:2.3b}\\
    J_{-}&=\left(-\partial_{x}-\frac{J_{3}}{x}+\frac{g}{J_{3}}\right)\mathrm{e}^{-i\theta}.\label{eq:2.3c}
  \end{align}
\end{subequations}
Here one may wonder about the meaning of $1/J_{3}$. The operator
$1/J_{3}$ would be defined as the spectral decomposition
$1/J_{3}=\sum_{j}(1/j)P_{j}$, where $P_{j}$ stands for the projection
operator onto the eigenspace of $J_{3}$ with eigenvalue $j$. This
definition would be well-defined unless the spectrum of $J_{3}$
contains $j=0$. An alternative way to give a meaning to $1/J_{3}$
would be the (formal) power series
$\frac{1}{J_{3}}=\frac{1}{\lambda}\frac{1}{1-(1-J_{3}/\lambda)}=\frac{1}{\lambda}\sum_{n=0}^{\infty}(1-\frac{J_{3}}{\lambda})^{n}$,
where $\lambda$ is an arbitrary constant. This expression would be
well-defined if the operator norm of $1-J_{3}/\lambda$ satisfies
$\|1-J_{3}/\lambda\|<1$. For the moment, however, we will proceed the
discussion at the formal level.

It is not difficult to show that the operators
\eqref{eq:2.3a}--\eqref{eq:2.3c} satisfy the following commutation
relations:
\begin{subequations}
  \begin{align}
    [J_{3},J_{\pm}]&=\pm J_{\pm},\label{eq:2.4a}\\
    [J_{+},J_{-}]&=-\frac{g^{2}}{J_{3}^{2}}+\frac{g^{2}}{(J_{3}-1)^{2}},\label{eq:2.4b}
  \end{align}
\end{subequations}
which follow from
$\mathrm{e}^{\mp i\theta}J_{3}\mathrm{e}^{\pm i\theta}=J_{3}\pm1$ or
$J_{3}\mathrm{e}^{\pm i\theta}=\mathrm{e}^{\pm i\theta}(J_{3}\pm1)$.
It is also easy to check that the invariant operator of this algebraic
system is given by
\begin{align}
  H
  &=J_{-}J_{+}-\frac{g^{2}}{J_{3}^{2}}\nonumber\\
  &=J_{+}J_{-}-\frac{g^{2}}{(J_{3}-1)^{2}}\nonumber\\
  &=-\partial_{x}^{2}+\frac{J_{3}(J_{3}-1)}{x^{2}}-\frac{2g}{x},\label{eq:2.5}
\end{align}
which commutes with $J_{\pm}$ and $J_{3}$.\footnote{The commutation
  relation $[H,J_{3}]=0$ is trivial. In order to prove
  $[H,J_{\pm}]=0$, one should first note that
  $HJ_{+}-J_{+}H=g^{2}(J_{+}\frac{1}{J_{3}^{2}}-\frac{1}{(J_{3}-1)^{2}}J_{+})$
  and
  $HJ_{-}-J_{-}H=g^{2}(J_{-}\frac{1}{(J_{3}-1)^{2}}-\frac{1}{J_{3}^{2}}J_{-})$,
  which follow from the first two lines of \eqref{eq:2.5}. Then by
  using \eqref{eq:2.3b}, \eqref{eq:2.3c}, and
  $\mathrm{e}^{-i\theta}\frac{1}{(J_{3}-1)^{2}}\mathrm{e}^{i\theta}=\frac{1}{J_{3}^{2}}$,
  one arrives at $[H,J_{\pm}]=0$.} Notice that if $g=0$ the
commutation relations \eqref{eq:2.4a} and \eqref{eq:2.4b} just
describe those for the Lie algebra $\mathfrak{iso}(2)$ of the
two-dimensional Euclidean group. In this case the invariant operator
$H$ is nothing but the Casimir operator of the Lie algebra
$\mathfrak{iso}(2)$.

\begin{figure}[t]
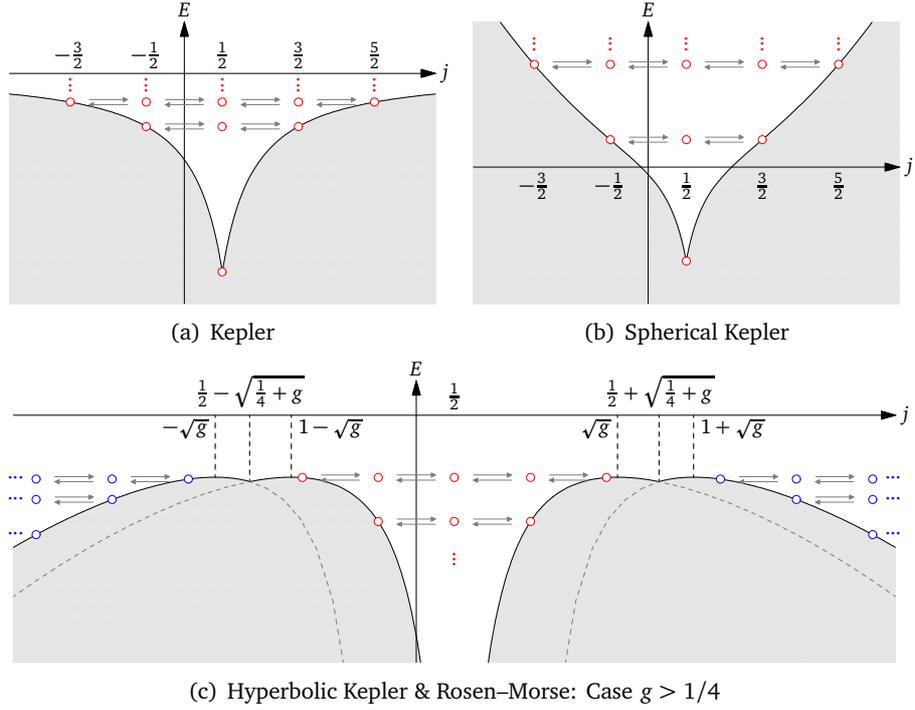

  \centering
  \subfigure[Kepler\label{figure:2a}]{\input{figure3a.eepic}}
  \subfigure[Spherical Kepler\label{figure:2b}]{\input{figure3b.eepic}}
  \subfigure[Hyperbolic Kepler \& Rosen--Morse: Case $g>1/4$\label{figure:2c}]{\input{figure3c.eepic}}
  \caption{Representations of the potential algebras. Gray shaded
    regions are the domains in which the squared norms
    $\|J_{\pm}|E,j\rangle\|^{2}$ become negative. Red circles
    represent the finite-dimensional representations, whereas blue
    circles represent the infinite-dimensional representations. Right
    and left arrows indicate the actions of ladder operators $J_{+}$
    and $J_{-}$, respectively.}
  \label{figure:2}
\end{figure}

Now, let $|E,j\rangle$ be a simultaneous eigenstate of $H$ and $J_{3}$
that satisfies the eigenvalue equations
\begin{subequations}
  \begin{align}
    H|E,j\rangle&=E|E,j\rangle,\label{eq:2.6a}\\
    J_{3}|E,j\rangle&=j|E,j\rangle,\label{eq:2.6b}
  \end{align}
\end{subequations}
and the normalization condition $\||E,j\rangle\|=1$. We wish to find
the possible values of $E$ and $j$. To this end, let us next consider
the states $J_{\pm}|E,j\rangle$.  As usual, the commutation relations
\eqref{eq:2.4a} lead
$J_{3}J_{\pm}|E,j\rangle=(j\pm1)J_{\pm}|E,j\rangle$, which implies
$J_{\pm}$ raise and lower the eigenvalue $j$ by $\pm1$:
\begin{align}
  J_{\pm}|E,j\rangle\propto|E,j\pm1\rangle.\label{eq:2.7}
\end{align}
Proportional coefficients are determined by calculating the norms
$\|J_{\pm}|E,j\rangle\|$. By using
$\|J_{\pm}|E,j\rangle\|^{2}=\langle E,j|J_{\mp}J_{\pm}|E,j\rangle$,
$J_{-}J_{+}=H+g^{2}/J_{3}^{2}$, and
$J_{+}J_{-}=H+g^{2}/(J_{3}-1)^{2}$, we get
\begin{subequations}
  \begin{align}
    \|J_{+}|E,j\rangle\|^{2}&=E+\frac{g^{2}}{j^{2}}\geq0,\label{eq:2.8a}\\
    \|J_{-}|E,j\rangle\|^{2}&=E+\frac{g^{2}}{(j-1)^{2}}\geq0.\label{eq:2.8b}
  \end{align}
\end{subequations}
These equations not only fix the proportional coefficients in
\eqref{eq:2.7} but also provide nontrivial constraints on $E$ and
$j$. In fact, together with the ladder equations \eqref{eq:2.7}, the
conditions \eqref{eq:2.8a} and \eqref{eq:2.8b} completely fix the
possible values of $E$ and $j$. To see this, let us consider a
negative-energy state $|E,j\rangle$ that corresponds to an arbitrary
point in the lower half of the $(E,j)$-plane. By applying the ladder
operators $J_{\pm}$ to the state $|E,j\rangle$ one can easily see that
such an arbitrary point eventually falls into the region in which the
squared norms become negative. See the figure below:
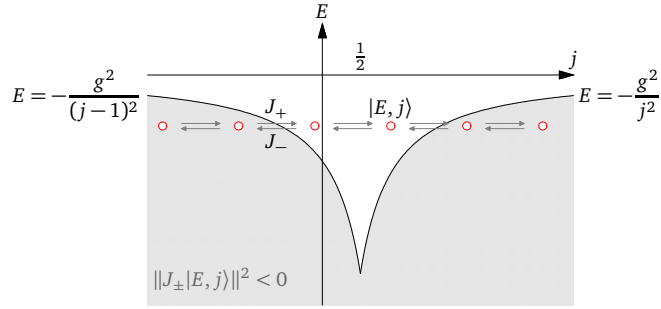
\begin{center}
  \input{figure2.eepic}
\end{center}
The only way to avoid this is to terminate the sequence
$\{\cdots,|E,j-1\rangle,|E,j\rangle,|E,j+1\rangle,\cdots\}$ from both
above and below. This is possible if and only if there exist both the
highest and lowest weight states $|E,j_{\text{max}}\rangle$ and
$|E,j_{\text{min}}\rangle$ in the sequence such that
$J_{+}|E,j_{\text{max}}\rangle=0=J_{-}|E,j_{\text{min}}\rangle$,
$-g^{2}/j_{\text{max}}^{2}=-g^{2}/(j_{\text{min}}-1)^{2}$,
$j_{\text{max}}-j_{\text{min}}\in\mathbb{Z}_{\geq0}$, and
$j_{\text{max}}\geq1/2$ and $j_{\text{min}}\leq1/2$. It is not
difficult to see that these conditions are fulfilled if and only if
the eigenvalue of the invariant operator takes the value
$E=-g^{2}/\nu^{2}$,
$\nu\in\{\tfrac{1}{2},1,\tfrac{3}{2},2,\cdots\}$. With this $\nu$ the
eigenvalues of $J_{3}$ take the values
$\{j_{\text{max}}=\nu,\nu-1,\cdots,2-\nu,j_{\text{min}}=1-\nu\}$. Note,
however, that if $\nu$ is an integer, the spectrum of $J_{3}$ contains
$j=0$ which makes the operator $1/J_{3}$ ill-defined. Thus we should
disregard this case. To summarize, the representation of the potential
algebra is specified by a half-integer
$\nu\in\{\tfrac{1}{2},\tfrac{3}{2},\cdots\}$ and the representation
space is spanned by the following $2\nu$ vectors:
\begin{align}
  \left\{|E,j\rangle: E=-\frac{g^{2}}{\nu^{2}}~~\&~~j\in\{\nu,\nu-1,\cdots,1-\nu\}\right\}.\label{eq:2.9}
\end{align}
These $2\nu$-dimensional representations are schematically depicted in
Figure \ref{figure:2a}.

Now it is straightforward to solve the original spectral problem of
the Kepler Hamiltonian \eqref{eq:1.1a}. To this end, let
$j\in\{\pm\tfrac{1}{2},\pm\tfrac{3}{2},\cdots\}$ be fixed. Since the
Hamiltonian is invariant under $j\to1-j$, without any loss of
generality we can focus on the case
$j\in\{\tfrac{1}{2},\tfrac{3}{2},\cdots\}$. Then the discrete energy
eigenvalues read
\begin{align}
  E_{n}=-\frac{g^{2}}{(j+n)^{2}},\quad n\in\{0,1,\cdots\}.\label{eq:2.10}
\end{align}
The energy eigenfunction $\psi_{E_{n},j}(x)$ that satisfies the
Schr\"{o}dinger equation
$H_{\text{Kepler}}\psi_{E_{n},j}=E_{n}\psi_{E_{n},j}$ can be
determined by the formula
$|E_{n},j\rangle\propto(J_{-})^{n}|E_{n},j+n\rangle$. Noting that
$|E,j\rangle$ corresponds to the function
$\psi_{E,j}(x)\mathrm{e}^{ij\theta}$ and $J_{-}$ is given by
$J_{-}=A_{-}(J_{3})\mathrm{e}^{-i\theta}$, we get the following
Rodrigues-like formula:
\begin{align}
  \psi_{E_{n},j}(x)\propto A_{-}(j)A_{-}(j+1)\cdots A_{-}(j+n-1)\psi_{E_{n},j+n}(x),\label{eq:2.11}
\end{align}
where $\psi_{E_{n},j+n}(x)$ is a solution to the first-order
differential equation $A_{+}(j+n)\psi_{E_{n},j+n}(x)=0$ and given by
$\psi_{E_{n},j+n}(x)\propto x^{j+n}\exp(-\frac{g}{j+n}x)$. All of
these exactly coincide with the well-known results.

In the rest of the note we would like to apply the same idea to the
spectral problem for the spherical Kepler, hyperbolic Kepler, and
Rosen--Morse Hamiltonians. We shall first introduce the potential
algebras, and then classify their representations, and then solve the
bound-state problems. As we will see below, the spherical Kepler
problem is rather straightforward but the hyperbolic Kepler and
Rosen--Morse potential problems are more intriguing and require
careful analysis.

\section{Spherical Kepler}
\label{section:3}
Let us next move on to the spherical Kepler problem
\cite{Schrodinger:1940xj,Infeld:1941,Stevenson:1941}, whose
Hamiltonian \eqref{eq:1.1b} is factorized as follows:
\begin{align}
  H_{\text{spherical Kepler}}=A_{-}(j)A_{+}(j)+j^{2}-\frac{g^{2}}{j^{2}},\label{eq:3.1}
\end{align}
where
\begin{align}
  A_{\pm}(j)=\pm\frac{d}{dx}-j\cot x+\frac{g}{j}.\label{eq:3.2}
\end{align}
Just as in the previous section, let us next introduce the following
operators:
\begin{subequations}
  \begin{align}
    J_{3}&=-i\partial_{\theta},\label{eq:3.3a}\\
    J_{+}&=\mathrm{e}^{i\theta}\left(\partial_{x}-\cot xJ_{3}+\frac{g}{J_{3}}\right),\label{eq:3.3b}\\
    J_{-}&=\left(-\partial_{x}-\cot xJ_{3}+\frac{g}{J_{3}}\right)\mathrm{e}^{-i\theta},\label{eq:3.3c}
  \end{align}
\end{subequations}
which satisfy the following commutation relations:
\begin{subequations}
  \begin{align}
    [J_{3},J_{\pm}]&=\pm J_{\pm},\label{eq:3.4a}\\
    [J_{+},J_{-}]&=J_{3}^{2}-\frac{g^{2}}{J_{3}^{2}}-(J_{3}-1)^{2}+\frac{g^{2}}{(J_{3}-1)^{2}}.\label{eq:3.4b}
  \end{align}
\end{subequations}
The invariant operator that commutes with $J_{3}$ and $J_{\pm}$ is
given by
\begin{align}
  H
  &=J_{-}J_{+}+J_{3}^{2}-\frac{g^{2}}{J_{3}^{2}}\nonumber\\
  &=J_{+}J_{-}+(J_{3}-1)^{2}-\frac{g^{2}}{(J_{3}-1)^{2}}\nonumber\\
  &=-\partial_{x}^{2}+\frac{J_{3}(J_{3}-1)}{\sin^{2}x}-2g\cot x.\label{eq:3.5}
\end{align}
It should be noted that, if $g=0$, Eqs.~\eqref{eq:3.4a} and
\eqref{eq:3.4b} reduce to the standard commutation relations for the
Lie algebra $\mathfrak{so}(3)$ under the appropriate shift
$J_{3}\to J_{3}+1/2$. In this case the invariant operator $H$ is
nothing but the Casimir operator of $\mathfrak{so}(3)$ and provides a
well-known example of interplay between shape invariance and Lie
algebra; see, e.g., the review \cite{Rasinariu:2007}.

Now, let $|E,j\rangle$ be a simultaneous eigenstate of $H$ and $J_{3}$
that satisfies the eigenvalue equations
\begin{subequations}
  \begin{align}
    H|E,j\rangle&=E|E,j\rangle,\label{eq:3.6a}\\
    J_{3}|E,j\rangle&=j|E,j\rangle,\label{eq:3.6b}
  \end{align}
\end{subequations}
as well as the normalization condition $\||E,j\rangle\|=1$. Then we
have the following conditions:
\begin{subequations}
  \begin{align}
    \|J_{+}|E,j\rangle\|^{2}&=E-j^{2}+\frac{g^{2}}{j^{2}}\geq0,\label{eq:3.7a}\\
    \|J_{-}|E,j\rangle\|^{2}&=E-(j-1)^{2}+\frac{g^{2}}{(j-1)^{2}}\geq0,\label{eq:3.7b}
  \end{align}
\end{subequations}
which, together with the ladder equations
$J_{\pm}|E,j\rangle\propto|E,j\pm1\rangle$, restrict the possible
values of $E$ and $j$. As discussed in the previous section, these
conditions are compatible with each other if and only if the
eigenvalue of the invariant operator takes the value
$E=\nu^{2}-g^{2}/\nu^{2}$,
$\nu\in\{\tfrac{1}{2},\tfrac{3}{2},\cdots\}$.  Now let
$\nu\in\{\tfrac{1}{2},\tfrac{3}{2},\cdots\}$ be fixed. Then the
representation space is spanned by the following $2\nu$ vectors:
\begin{align}
  \left\{|E,j\rangle: E=\nu^{2}-\frac{g^{2}}{\nu^{2}}~~\&~~j\in\{\nu,\nu-1,\cdots,1-\nu\}\right\}.\label{eq:3.8}
\end{align}
These $2\nu$-dimensional representations are schematically depicted in
Figure \ref{figure:2b}.

Now it is easy to find the spectrum of the original Hamiltonian
\eqref{eq:1.1b}. For fixed $j\in\{\tfrac{1}{2},\tfrac{3}{2},\cdots\}$
the energy eigenvalues and eigenfunctions read
\begin{align}
  E_{n}=(j+n)^{2}-\frac{g^{2}}{(j+n)^{2}},\quad n\in\{0,1,\cdots\},\label{eq:3.9}
\end{align}
and
\begin{align}
  \psi_{E_{n},j}(x)\propto A_{-}(j)A_{-}(j+1)\cdots A_{-}(j+n-1)\psi_{E_{n},j+n}(x),\label{eq:3.10}
\end{align}
where
$\psi_{E_{n},j+n}(x)\propto (\sin x)^{j+n}\exp(-\frac{g}{j+n}x)$. We
note that Eqs.~\eqref{eq:3.9} and \eqref{eq:3.10} are consistent with
the known results
\cite{Schrodinger:1940xj,Infeld:1941,Stevenson:1941}.

\section{Hyperbolic Kepler \& Rosen--Morse}
\label{section:4}
Let us finally move on to the study of potential algebras for the
hyperbolic Kepler and Rosen--Morse Hamiltonians. We shall see that the
bound-state spectra of these problems correspond to two distinct
representations of a single algebraic system.
\subsection{Hyperbolic Kepler}
\label{section:4.1}
The Hamiltonian \eqref{eq:1.1c} for the hyperbolic Kepler problem
\cite{Manning:1933,Infeld:1945} can be factorized as follows:
\begin{align}
  H_{\text{hyperbolic Kepler}}=A_{-}(j)A_{+}(j)-j^{2}-\frac{g^{2}}{j^{2}},\label{eq:4.1}
\end{align}
where
\begin{align}
  A_{\pm}(j)=\pm\frac{d}{dx}-j\coth x+\frac{g}{j}.\label{eq:4.2}
\end{align}
We then introduce the following operators:
\begin{subequations}
  \begin{align}
    J_{3}&=-i\partial_{\theta},\label{eq:4.3a}\\
    J_{+}&=\mathrm{e}^{i\theta}\left(\partial_{x}-\coth xJ_{3}+\frac{g}{J_{3}}\right),\label{eq:4.3b}\\
    J_{-}&=\left(-\partial_{x}-\coth xJ_{3}+\frac{g}{J_{3}}\right)\mathrm{e}^{-i\theta},\label{eq:4.3c}
  \end{align}
\end{subequations}
which satisfy the following commutation relations:
\begin{subequations}
  \begin{align}
    [J_{3},J_{\pm}]&=\pm J_{\pm},\label{eq:4.4a}\\
    [J_{+},J_{-}]&=-J_{3}^{2}-\frac{g^{2}}{J_{3}^{2}}+(J_{3}-1)^{2}+\frac{g^{2}}{(J_{3}-1)^{2}}.\label{eq:4.4b}
  \end{align}
\end{subequations}
The invariant operator is given by
\begin{align}
  H
  &=J_{-}J_{+}-J_{3}^{2}-\frac{g^{2}}{J_{3}^{2}}\nonumber\\
  &=J_{+}J_{-}-(J_{3}-1)^{2}-\frac{g^{2}}{(J_{3}-1)^{2}}\nonumber\\
  &=-\partial_{x}^{2}+\frac{J_{3}(J_{3}-1)}{\sinh^{2}x}-2g\coth x.\label{eq:4.5}
\end{align}
We note that, if $g=0$, Eqs.~\eqref{eq:4.4a} and \eqref{eq:4.4b}
reduce to the standard commutation relations for the Lie algebra
$\mathfrak{so}(2,1)$ under the shift $J_{3}\to J_{3}+1/2$. In other
words, the operators \eqref{eq:4.3a}--\eqref{eq:4.3c} provide one of
differential realizations of $\mathfrak{so}(2,1)$ if $g=0$ and
$J_{3}\to J_{3}+1/2$. Unfortunately, however, this Lie-algebraic
structure is less useful in the present problem because the invariant
operator \eqref{eq:4.5} does not contain discrete eigenvalues if $g=0$
and $J_{3}$ has real eigenvalues. As we will see shortly, however,
this situation gets changed if $g$ is non-vanishing.

Now, let $|E,j\rangle$ be a simultaneous eigenstate of $H$ and
$J_{3}$:
\begin{subequations}
  \begin{align}
    H|E,j\rangle&=E|E,j\rangle,\label{eq:4.6a}\\
    J_{3}|E,j\rangle&=j|E,j\rangle.\label{eq:4.6b}
  \end{align}
\end{subequations}
Then, under the normalization condition $\||E,j\rangle\|=1$, the
squared norms $\|J_{\pm}|E,j\rangle\|^{2}$ are evaluated as follows:
\begin{subequations}
  \begin{align}
    \|J_{+}|E,j\rangle\|^{2}&=E+j^{2}+\frac{g^{2}}{j^{2}}\geq0,\label{eq:4.7a}\\
    \|J_{-}|E,j\rangle\|^{2}&=E+(j-1)^{2}+\frac{g^{2}}{(j-1)^{2}}\geq0.\label{eq:4.7b}
  \end{align}
\end{subequations}
These conditions are enough to classify representations. In contrast
to the previous two examples, there are several nontrivial
representations depending on the range of $j$. For $g>1/4$, we have
the following three distinct representations (see Figure
\ref{figure:2c}):
\begin{itemize}
\item \textbf{\boldmath Case $j\in(-\infty,-\sqrt{g})$:
    Infinite-dimensional representation.} Let
  $\nu\in(-\infty,-\sqrt{g})$ be fixed. Then the representation space
  is spanned by the following infinitely many vectors:
  \begin{align}
    \left\{|E,j\rangle: E=-\nu^{2}-\frac{g^{2}}{\nu^{2}}~~\&~~j\in\{\nu,\nu-1,\cdots\}\right\}.\label{eq:4.8}
  \end{align}
  We emphasize that in this case the parameter
  $\nu\in(-\infty,-\sqrt{g})$ is not necessarily restricted to an
  integer or half-integer. This is a one-parameter family of
  infinite-dimensional representation of the algebraic system
  $\{J_{3},J_{+},J_{-}\}$.
\item \textbf{\boldmath Case $j\in(1-\sqrt{g},\sqrt{g})$:
    Finite-dimensional representation.} Let
  $\nu\in\{\tfrac{1}{2},\tfrac{3}{2},\cdots,\nu_{\text{max}}\}$ be
  fixed, where $\nu_{\text{max}}$ is the maximal half-integer smaller
  than $\sqrt{g}$; i.e.,
  $\nu_{\text{max}}=\max\{\nu\in\tfrac{1}{2}\mathbb{N}:\nu<\sqrt{g}\}$. Then
  the representation space is spanned by the following $2\nu$ vectors:
  \begin{align}
    \left\{|E,j\rangle: E=-\nu^{2}-\frac{g^{2}}{\nu^{2}}~~\&~~j\in\{\nu,\nu-1,\cdots,1-\nu\}\right\}.\label{eq:4.9}
  \end{align}
  This is a $2\nu$-dimensional representation of the algebraic system
  $\{J_{3},J_{+},J_{-}\}$.
\item \textbf{\boldmath Case $j\in(1+\sqrt{g},\infty)$:
    Infinite-dimensional representation.} Let
  $\nu\in(1+\sqrt{g},\infty)$ be fixed. Then the representation space
  is spanned by the following infinitely many vectors:
  \begin{align}
    \left\{|E,j\rangle: E=-(\nu-1)^{2}-\frac{g^{2}}{(\nu-1)^{2}}~~\&~~j\in\{\nu,\nu+1,\cdots\}\right\}.\label{eq:4.10}
  \end{align}
  Note that $\nu\in(1+\sqrt{g},\infty)$ is a continuous parameter and
  is not necessarily be an integer or half-integer. This is another
  one-parameter family of infinite-dimensional representation of the
  algebraic system $\{J_{3},J_{+},J_{-}\}$.
\end{itemize}
One may notice that the region
$[-\sqrt{g},1-\sqrt{g}]\cup[\sqrt{g},1+\sqrt{g}]$ is excluded in the
above classification. This is because there is no bound state in this
region for both the hyperbolic Kepler and Rosen--Morse potential
problems. We note that the finite-dimensional representation
\eqref{eq:4.9} disappears for $g\leq1/4$, whereas the
infinite-dimensional representations \eqref{eq:4.8} and
\eqref{eq:4.10} remain present for $g\leq1/4$.

Now we have classified the representations of the potential
algebra. The next task we have to do is to understand which
representations are realized in the hyperbolic Kepler problem. To see
this, let us consider the potential
$V(x)=j(j-1)/\sinh^{2}x-2g\coth x$. In order to have a bound state, it
is necessary that $V(x)$ has a minimum on the half line.\footnote{This
  is, of course, not sufficient condition.} This is achieved if and
only if $j$ is in the range
$(\tfrac{1}{2}-\sqrt{g+\tfrac{1}{4}},\tfrac{1}{2}+\sqrt{g+\tfrac{1}{4}})$,
which includes $(1-\sqrt{g},\sqrt{g})$; see Figure
\ref{figure:2c}. Hence the bound state spectrum should be related to
the finite-dimensional representation \eqref{eq:4.9}.

Now it is easy to solve the original eigenvalue problem
$H_{\text{hyperbolic Kepler}}\psi_{E_{n},j}=E_{n}\psi_{E_{n},j}$ for
the hyperbolic Kepler Hamiltonian. For fixed
$j\in\{\tfrac{1}{2},\tfrac{3}{2},\cdots,\nu_{\text{max}}\}$, the
energy eigenvalues and eigenfunctions are given by
\begin{align}
  E_{n}=-(j+n)^{2}-\frac{g^{2}}{(j+n)^{2}},\quad n\in\{0,1,\cdots,N\},\label{eq:4.11}
\end{align}
and
\begin{align}
  \psi_{E_{n},j}(x)\propto A_{-}(j)A_{-}(j+1)\cdots A_{-}(j+n-1)\psi_{E_{n},j+n}(x),\label{eq:4.12}
\end{align}
where
$N=\max\{n\in\mathbb{Z}_{\geq0}:j+n<\sqrt{g}\}=\nu_{\text{max}}-j$ and
$\psi_{E_{n},j+n}(x)\propto(\sinh
x)^{j+n}\exp(-\frac{g}{j+n}x)$. Notice that these results are
consistent with the known results \cite{Infeld:1945}.

Before closing this subsection it is worthwhile to comment on the case
$g\leq1/4$. As mentioned before, the finite-dimensional representation
\eqref{eq:4.9} disappears for $g\leq1/4$. However, new
finite-dimensional representations appear in this case. The relevant
one is the following one-dimensional representation spanned by a
single vector:
\begin{align}
  \left\{|E,j\rangle: E=-j^{2}-\frac{g^{2}}{j^{2}}~~\&~~j=\frac{1}{2}-\sqrt{\frac{1}{4}-g}\right\},
\end{align}
where $g\in(0,1/4)$. Notice that this $j$ is one of the solutions to
the condition $-j^{2}-g^{2}/j^{2}=-(j-1)^{2}-g^{2}/(j-1)^{2}$. Now one
can easily check that this state vector satisfies
$J_{\pm}|E,j\rangle=0$. It is also easy to see that, for
$g\in(0,1/4)$, $j=1/2-\sqrt{1/4-g}$ satisfies the condition
$j<\sqrt{g}$, which is the necessary condition for the ground-state
wavefunction to be normalizable. The point is that, just as in the
case $g>1/4$, $j$ must be quantized in a particular manner in this
representation theoretic approach.

\subsection{Rosen--Morse}
\label{section:4.2}
Let us finally move on to the bound-state problem of the Rosen--Morse
Hamiltonian \cite{Eckart:1930zza,Rosen:1932}. First, the Hamiltonian
\eqref{eq:1.1d} is factorized as follows:
\begin{align}
  H_{\text{Rosen--Morse}}=A_{-}(j)A_{+}(j)-j^{2}-\frac{g^{2}}{j^{2}},\label{eq:4.13}
\end{align}
where
\begin{align}
  A_{\pm}(j)=\pm\frac{d}{dx}-j\tanh x+\frac{g}{j}.\label{eq:4.14}
\end{align}
Let us then introduce the following operators:
\begin{subequations}
  \begin{align}
    J_{3}&=-i\partial_{\theta},\label{eq:4.15a}\\
    J_{+}&=\mathrm{e}^{i\theta}\left(\partial_{x}-\tanh xJ_{3}+\frac{g}{J_{3}}\right),\label{eq:4.15b}\\
    J_{-}&=\left(-\partial_{x}-\tanh xJ_{3}+\frac{g}{J_{3}}\right)\mathrm{e}^{-i\theta},\label{eq:4.15c}
  \end{align}
\end{subequations}
which satisfy the commutation relations:
\begin{subequations}
  \begin{align}
    [J_{3},J_{\pm}]&=\pm J_{\pm},\label{eq:4.16a}\\
    [J_{+},J_{-}]&=-J_{3}^{2}-\frac{g^{2}}{J_{3}^{2}}+(J_{3}-1)^{2}+\frac{g^{2}}{(J_{3}-1)^{2}}.\label{eq:4.16b}
  \end{align}
\end{subequations}
The invariant operator is
\begin{align}
  H
  &=J_{-}J_{+}-J_{3}^{2}-\frac{g^{2}}{J_{3}^{2}}\nonumber\\
  &=J_{+}J_{-}-(J_{3}-1)^{2}-\frac{g^{2}}{(J_{3}-1)^{2}}\nonumber\\
  &=-\partial_{x}^{2}-\frac{J_{3}(J_{3}-1)}{\cosh^{2}x}-2g\tanh x.\label{eq:4.17}
\end{align}
Note that the commutation relations \eqref{eq:4.16a} and
\eqref{eq:4.16b} are exactly the same as those for the hyperbolic
Kepler problem. Hence the bound-state spectrum should be related to
the representations classified in the previous subsection. To
understand which representations are realized, let us study the
minimum of the potential $V(x)=-j(j-1)/\cosh^{2}x-2g\tanh x$. Thanks
to the symmetry $j\to1-j$, without any loss of generality we can focus
on the case $j\geq1/2$. It is then easy to see that the potential has
a minimum if $j$ is in the range
$(\tfrac{1}{2}+\sqrt{g+\tfrac{1}{4}},\infty)$, which contains the
region $(1+\sqrt{g},\infty)$; see Figure \ref{figure:2c}. Hence, in
contrast to the previous case, the bound-state problem for the
Rosen--Morse Hamiltonian should be related to the infinite-dimensional
representation \eqref{eq:4.10}.

Now it is easy to find the energy eigenvalue of the original
Hamiltonian. Let $j\in(1+\sqrt{g},\infty)$ be fixed. Then the energy
eigenvalues and eigenfunctions read
\begin{align}
  E_{n}=-(j-n-1)^{2}-\frac{g^{2}}{(j-n-1)^{2}},\quad n\in\{0,1,\cdots,N\},\label{eq:4.18}
\end{align}
and
\begin{align}
  \psi_{E_{n},j}(x)\propto A_{+}(j-1)A_{+}(j-2)\cdots A_{+}(j-n)\psi_{E_{n},j-n}(x),\label{eq:4.19}
\end{align}
where $N=\max\{n\in\mathbb{Z}_{\geq0}:1+\sqrt{g}<j-n\}$ and
$\psi_{E_{n},j-n}(x)\propto(\cosh
x)^{-j+n+1}\exp(\tfrac{g}{j-n-1}x)$. Notice that Eqs.~\eqref{eq:4.18}
and \eqref{eq:4.19} are consistent with the known results
\cite{Rosen:1932}.

\section{Conclusions}
\label{section:5}
In this note we have revisited the bound-state problems for the
Kepler, spherical Kepler, hyperbolic Kepler, and Rosen--Morse
Hamiltonians, all of which have not been solved before in terms of
potential algebra. We have introduced three nonlinear algebraic
systems and solved the problems by means of representation theory. We
have seen that the discrete energy spectra can be obtained just from
the four conditions: $J_{\pm}|E,j\rangle\propto|E,j\pm1\rangle$ and
$\|J_{\pm}|E,j\rangle\|^{2}\geq0$. These conditions correctly
reproduce the known results in a purely algebraic fashion. The price
to pay, however, is that in this approach $j$ must be a half-integer
(except for the Rosen--Morse potential problem and the hyperbolic
Kepler problem in the domain $g\in(0,1/4)$), otherwise there arise
inconsistencies. This is a weakness of this representation theoretic
approach.

\bibliographystyle{utphys}%
\bibliography{bibliography}%

\providecommand{\href}[2]{#2}\begingroup\raggedright\begin{thebibliography}{10}

\bibitem{Schrodinger:1940xj}
E.~Schr{\"o}dinger, ``{A Method of Determining Quantum-Mechanical Eigenvalues
  and Eigenfunctions},''
\href{http://www.jstor.org/stable/20490744}{{\em Proc. Roy. Irish Acad. (Sect. A)} {\bfseries 46} (1940) 9--16}.

\bibitem{Infeld:1941}
L.~Infeld, ``{On a New Treatment of Some Eigenvalue Problems},''
  \href{http://dx.doi.org/10.1103/PhysRev.59.737}{{\em Phys. Rev.} {\bfseries
  59} (1941) 737--747}.

\bibitem{Stevenson:1941}
A.~F. Stevenson, ``{Note on the ``Kepler Problem'' in a Spherical Space, and
  the Factorization Method of Solving Eigenvalue Problems},''
  \href{http://dx.doi.org/10.1103/PhysRev.59.842}{{\em Phys. Rev.} {\bfseries
  59} (1941) 842--843}.

\bibitem{Manning:1933}
M.~F. Manning and N.~Rosen, ``{A Potential Function for the Vibrations of
  Diatomic Molecule},'' \href{http://dx.doi.org/10.1103/PhysRev.44.951}{{\em
  Phys. Rev.} {\bfseries 44} (1933) 951--954}.

\bibitem{Infeld:1945}
L.~Infeld and A.~Schild, ``{A Note on the Kepler Problem in a Space of Constant
  Negative Curvature},'' \href{http://dx.doi.org/10.1103/PhysRev.67.121}{{\em
  Phys. Rev.} {\bfseries 67} (1945) 121--122}.

\bibitem{Eckart:1930zza}
C.~Eckart, ``{The Penetration of a Potential Barrier by Electrons},''
\href{http://dx.doi.org/10.1103/PhysRev.35.1303}{{\em Phys. Rev.} {\bfseries
  35} (1930) 1303--1309}.

\bibitem{Rosen:1932}
N.~Rosen and P.~M. Morse, ``{On the Vibrations of Polyatomic Molecules},''
  \href{http://dx.doi.org/10.1103/PhysRev.42.210}{{\em Phys. Rev.} {\bfseries
  42} (1932) 210--217}.

\bibitem{Infeld:1951mw}
L.~Infeld and T.~E. Hull, ``{The Factorization Method},''
\href{http://dx.doi.org/10.1103/RevModPhys.23.21}{{\em Rev. Mod. Phys.}
  {\bfseries 23} (1951) 21--68}.

\bibitem{Cooper:1994eh}
F.~Cooper, A.~Khare, and U.~Sukhatme, ``{Supersymmetry and quantum
  mechanics},'' \href{http://dx.doi.org/10.1016/0370-1573(94)00080-M}{{\em
  Phys. Rept.} {\bfseries 251} (1995) 267--385},
\href{http://arxiv.org/abs/hep-th/9405029}{{\ttfamily arXiv:hep-th/9405029
  [hep-th]}}.

\bibitem{Houri:2017xtq}
T.~Houri, M.~Sakamoto, and K.~Tatsumi, ``{Spectral intertwining relations in
  exactly solvable quantum-mechanical systems},''
  \href{http://dx.doi.org/10.1093/ptep/ptx074}{{\em PTEP} {\bfseries 2017}
  (2017) 063A01},
\href{http://arxiv.org/abs/1701.04307}{{\ttfamily arXiv:1701.04307
  [quant-ph]}}.

\bibitem{Gangopadhyaya:1998ccj}
A.~Gangopadhyaya, J.~V. Mallow, and U.~P. Sukhatme, ``{Translational shape
  invariance and the inherent potential algebra},''
\href{http://dx.doi.org/10.1103/PhysRevA.58.4287}{{\em Phys. Rev.} {\bfseries
  A58} (1998) 4287--4292}.

\bibitem{Rasinariu:2007}
C.~Rasinariu, J.~V. Mallow, and A.~Gangopadhyaya, ``{Exactly solvable problems
  of quantum mechanics and their spectrum generating algebras: A review},''
  \href{http://dx.doi.org/10.2478/s11534-007-0001-1}{{\em Central Eur. J.
  Phys.} {\bfseries 5} (2007) 111--134}.

\bibitem{Bougie:2012}
J.~Bougie, A.~Gangopadhyaya, J.~Mallow, and C.~Rasinariu, ``{Supersymmetric
  Quantum Mechanics and Solvable Models},''
  \href{http://dx.doi.org/10.3390/sym4030452}{{\em Symmetry} {\bfseries 4}
  (2012) 452--473}.

\bibitem{Balantekin:1997mg}
A.~B. Balantekin, ``{Algebraic approach to shape invariance},''
  \href{http://dx.doi.org/10.1103/PhysRevA.57.4188}{{\em Phys. Rev.} {\bfseries
  A57} (1998) 4188--4191},
\href{http://arxiv.org/abs/quant-ph/9712018}{{\ttfamily arXiv:quant-ph/9712018
  [quant-ph]}}.

\end{thebibliography}\endgroup
\end{document}